\documentclass[letterpaper, 10 pt, conference]{IEEEtran}

\usepackage[mathscr]{eucal}
\usepackage[cmex10]{amsmath}
\usepackage{epsfig,epsf,psfrag}
\usepackage{amssymb,amsmath,amsthm,amsfonts,latexsym}
\usepackage{amsmath,graphicx,bm,xcolor,url,overpic}
\usepackage{fixltx2e}
\usepackage{array}
\usepackage{verbatim}
\usepackage{bm}
\usepackage{algorithmic}
\usepackage{algorithm}
\usepackage{verbatim}
\usepackage{textcomp}
\usepackage{mathrsfs}
\usepackage{epstopdf}

\newcommand{\openone}{\leavevmode\hbox{\small1\normalsize\kern-.33em1}} 

\catcode`~=11 \def\UrlSpecials{\do\~{\kern -.15em\lower .7ex\hbox{~}\kern .04em}} \catcode`~=13 

\allowdisplaybreaks[4]

\newcommand{\calR}{\mathcal{R}}
\newcommand{\calS}{\mathcal{S}}

\newcommand{\calU}{\mathcal{U}}

\newcommand{\calX}{\mathcal{X}}

\DeclareMathAlphabet{\mathbsf}{OT1}{cmss}{bx}{n}
\DeclareMathAlphabet{\mathssf}{OT1}{cmss}{m}{sl}

\DeclareSymbolFont{bsfletters}{OT1}{cmss}{bx}{n}  
\DeclareSymbolFont{ssfletters}{OT1}{cmss}{m}{n}
\DeclareMathSymbol{\bsfGamma}{0}{bsfletters}{'000}
\DeclareMathSymbol{\ssfGamma}{0}{ssfletters}{'000}
\DeclareMathSymbol{\bsfDelta}{0}{bsfletters}{'001}
\DeclareMathSymbol{\ssfDelta}{0}{ssfletters}{'001}
\DeclareMathSymbol{\bsfTheta}{0}{bsfletters}{'002}
\DeclareMathSymbol{\ssfTheta}{0}{ssfletters}{'002}
\DeclareMathSymbol{\bsfLambda}{0}{bsfletters}{'003}
\DeclareMathSymbol{\ssfLambda}{0}{ssfletters}{'003}
\DeclareMathSymbol{\bsfXi}{0}{bsfletters}{'004}
\DeclareMathSymbol{\ssfXi}{0}{ssfletters}{'004}
\DeclareMathSymbol{\bsfPi}{0}{bsfletters}{'005}
\DeclareMathSymbol{\ssfPi}{0}{ssfletters}{'005}
\DeclareMathSymbol{\bsfSigma}{0}{bsfletters}{'006}
\DeclareMathSymbol{\ssfSigma}{0}{ssfletters}{'006}
\DeclareMathSymbol{\bsfUpsilon}{0}{bsfletters}{'007}
\DeclareMathSymbol{\ssfUpsilon}{0}{ssfletters}{'007}
\DeclareMathSymbol{\bsfPhi}{0}{bsfletters}{'010}
\DeclareMathSymbol{\ssfPhi}{0}{ssfletters}{'010}
\DeclareMathSymbol{\bsfPsi}{0}{bsfletters}{'011}
\DeclareMathSymbol{\ssfPsi}{0}{ssfletters}{'011}
\DeclareMathSymbol{\bsfOmega}{0}{bsfletters}{'012}
\DeclareMathSymbol{\ssfOmega}{0}{ssfletters}{'012}

\newcommand{\hatX}{\hat{X}}

\newtheorem{theorem}{Theorem}

\newtheorem{definition}{Definition}

\usepackage{amssymb}
\usepackage{amsmath}
\usepackage{booktabs}
\usepackage{amsfonts}
\usepackage{amsthm}
\usepackage{xfrac}
\usepackage{color}
\usepackage{array}
\usepackage{graphicx}
\usepackage[font=footnotesize]{caption}
\usepackage[font=footnotesize]{subcaption}
\usepackage{cite}
\usepackage{cases}
\usepackage{wasysym}
\usepackage{dsfont}
\usepackage{enumitem}
\usepackage[makeroom]{cancel}
\DeclareMathOperator*{\ld}{ld}

\newcommand\blfootnote[1]{%
	\begingroup
	\renewcommand\thefootnote{}\footnote{#1}%
	\addtocounter{footnote}{-1}%
	\endgroup
}

\begin{document}

\title{\vspace{6.3mm} On The Binary Lossless Many-Help-One Problem\\ with Independently Degraded Helpers}

\author{\IEEEauthorblockN{
		Albrecht Wolf\IEEEauthorrefmark{1},
		Diana Cristina Gonz\'{a}lez\IEEEauthorrefmark{2},
		Meik D{\"o}rpinghaus\IEEEauthorrefmark{1},\\
		Jos\'{e} C\^{a}ndido Silveira Santos Filho\IEEEauthorrefmark{2},
		and Gerhard Fettweis\IEEEauthorrefmark{1}\\
	}
	\IEEEauthorblockA{
		\IEEEauthorrefmark{1}Vodafone Chair Mobile Communications Systems, Technische Universit\"{a}t Dresden, Germany\\
		Email: \{albrecht.wolf, meik.doerpinghaus, gerhard.fettweis\}@tu-dresden.de.\\
		\IEEEauthorrefmark{2}Department of Communications, School of Electrical and Computer Engineering, University of Campinas, Brazil\\
		E-mail: \{dianigon, candido\}@decom.fee.unicamp.br.
	}
}


\maketitle

\begin{abstract}
Although the rate region for the lossless many-help-one problem with independently degraded helpers is already ``solved", its solution is given in terms of a convex closure over a set of auxiliary random variables. Thus, for any such a problem in particular, an optimization over the set of auxiliary random variables is required to truly solve the rate region. Providing the solution is surprisingly difficult even for an example as basic as binary sources. In this work, we derive a simple and tight inner bound on the rate region's lower boundary for the lossless many-help-one problem with independently degraded helpers when specialized to sources that are binary, uniformly distributed, and interrelated through symmetric channels. This scenario finds important applications in emerging cooperative communication schemes in which the direct-link transmission is assisted via multiple lossy relaying links.  Numerical results indicate that the derived inner bound proves increasingly tight as the helpers become more degraded.\blfootnote{This work has been submitted to the IEEE for possible publication. Copyright may be transferred without notice, after which this version may no longer
	be accessible.} 
\end{abstract}

%


\section{Introduction}
The idea that a decoder wishes to reproduce a primary source ($X_0$) with the help of an auxiliary source ($X_1$), introduced by Wyner \cite{Wyner1975}, Ahlswede, and K{\"o}rner \cite{Ahlswede1975}, can be intuitively extended to an arbitrary number of auxiliary sources $(X_1,...,X_L)$ a.k.a. helpers. Finding the rate region of such a system defines the so-called many-help-one problem. This problem has been recognized as a highly challenging one, and only a few particular solutions are known to date. K{\"o}rner and Marton \cite{Korner1979} addressed a two-help-one problem where the primary source is a modulo-two sum of correlated binary auxiliary sources. Gelfand and Pinsker \cite{gel1979} determined the rate region when the auxiliary sources are discrete and conditionally independent if the primary source is given. Motivated by the Gelfand-Pinsker result, Oohama \cite{Oohama2005} determined the rate-distortion region for the same setup but Gaussian sources. Tavildar \cite{Tavildar2010} derived the rate-distortion region for Gaussian sources with a correlation model following a tree-like structure. For other works on the many-help-one problem, see \cite{Oohama2008} and the references therein.

While the characterization given by Gelfand and Pinsker \cite{gel1979} is elegant and quite general, it presents a practical disadvantage: the solution relies on auxiliary random variables (RVs) whose statistics are unknown in advance. Thus, the numerical characterization of the region of achievable rates for any particular joint distribution of ($X_0,X_1,...,X_L$) requires an optimization over all admissible conditional distributions for the auxiliary RVs $(U_1,...,U_L)$. Gelfand and Pinsker \cite{gel1979} showed that the rate region remains unchanged if the alphabet size of the auxiliary RVs is bounded by $|\calU_l|\leq |\calX_l|+(L+1)2^{L-1}+1$. However, with an increasing number of helpers the bound on the alphabet size increases, and so does the complexity of the optimization problem. Jana~\cite{Jana2009} showed that the cardinality of the auxiliary RVs can be tightly bounded by $|\calU_l| \leq |\calX_l|$ for a broad class of multiterminal source coding problems, including the many-help-one problem. But still, the optimization problem remains surprisingly challenging, even for binary sources.

In \cite{Gu2007}, the one-help-one problem (a.k.a. source coding with coded side information) was considered with binary sources which are related through a binary symmetric channel (BSC).  It was then shown that the rate region is achieved if and only if the auxiliary RV and the helper are related through a BSC as well.

In this work, we investigate the many-help-one problem when specialized to source and helpers that are binary, uniformly distributed, and interrelated through BSCs. Motivated by the results in \cite{Gu2007}, we assume the helpers and auxiliary RVs are also interrelated through BSCs, thereby deriving a simple and tight inner bound on the rate region's lower boundary for the investigated problem. The more degraded the helpers, the tighter the inner bound, as indicated from our numerical examples.

\section{Background}
\begin{figure}[t]
	\centering
		\includegraphics[width=\linewidth]{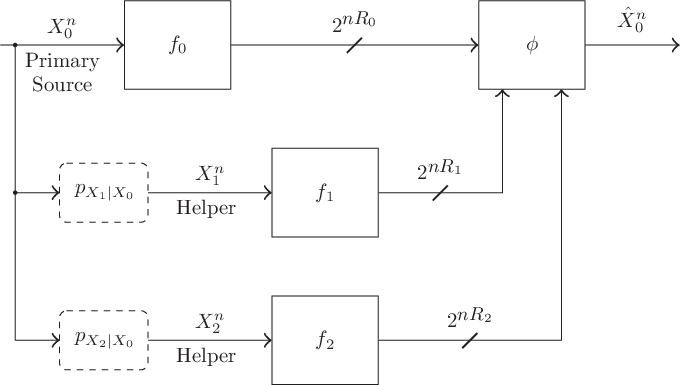}
		\caption{System model for the lossless many-help-one problem with two independently degraded helpers (i.e., $L=2$). We are interested in the rate region for sources that are binary, uniformly distributed, and interrelated through symmetric channels.}
		\label{fig:SystemModel}
\end{figure}

\subsection{Notation}
Random variables and their realizations are denoted in capital (e.g.,\ $X$) and lowercase (e.g.,\ $x$) letters, respectively. All sets (e.g., alphabets of random variables) are denoted in calligraphic letter (e.g.,\ $\mathcal{X}$). Also, $X^n:=(X_1,\ldots,X_n)$ denotes a random vector of length $n$.

Given any two integers $a,b\in\mathds{N}$, we use $[a:b]$ to denote the inclusive collection of all integers between $a$ and $b$, i.e., $[a:b]:=\{c:c\in\mathds{N}~,a\leq c\leq b\}$. Furthermore, we use $[a]$ as a compact notation to $[1:a]$, for any integer $a$.

Finally, we denote the probability of an event $\mathcal{E}$ as $\text{Pr}\{\mathcal{E}\}$, the mutual information as $I(\cdot;\cdot)$, the entropy as $H(\cdot)$, the binary logarithm as $\ld(\cdot)$, the binary entropy function as $h(p)=-p \ld (p) - (1-p) \ld (1-p)$, the binary convolution as $a_1 \ast a_2=a_1 (1-a_2) + (1-a_1) a_2$, and the multivariate binary convolution as $a_1 \ast ... \ast a_N =  a_1 \ast( ...\ast (a_{N-1} \ast a_N)...)$, which is a \emph{cascaded} binary convolution.
 
\subsection{Multiterminal Source Coding and Rate Region}
Assume that we are given $(L+1)$ memoryless sources, where one of these sources is the primary source, while the $L$ remaining ones are helpers. The primary source is generated i.i.d. according to a distribution defined on the finite alphabet $\calX_0$ of size $|\calX_0|<\infty$. The helpers are conditionally independent given the primary source on the finite alphabet $\calX_l$ of size $|\calX_l|<\infty$ for $l \in [L]$. Hereafter, this case is referred to as the CI condition. The joint probability mass function (pmf) of  $\{X_l\}_{l\in [0:L]}$ satisfies
\begin{align}
& p_{X_0X_1...X_L}(x_0,x_1,...,x_L) \nonumber \\
\label{eq:pmf_cond_dep}
& \qquad = p_{X_0}(x_0) \prod\limits_{l \in [L]} p_{X_l|X_0}(x_l | x_0).
\end{align}
Fig.~\ref{fig:SystemModel} illustrates the lossless many-help-one system model. 
\begin{definition}
\label{def:code}
An $(n,M_0,M_1,...,M_L)$-code consists of 
\begin{itemize}
	\item $(L+1)$ encoders
	\begin{align}
	f_l:\calX_l^n\to [M_l],~\forall~l\in[0:L],~and
	\end{align}
	\item a decoder
	\begin{align}
	\phi: \prod_{l\in[0:L]}[M_l]\to\calX_0^n.
	\end{align}
\end{itemize}
\end{definition}
Given an $(n,M_0,M_1,...,M_L)$-code, the primary source estimate can be expressed as 
\begin{align}
	\label{eq:reproduced_source}
	\hatX_0^n=\phi(\{f_l(X_l^n)\}_{l\in [0:L]})
\end{align} 
and, as criterion of fidelity of reproduction of sequence $X_0^n$, we will use the maximum error probability per source symbol:
\begin{align}
	p_{\max}:=\max_{i\in[n]}\Pr\{X_{0,i}\neq \hatX_{0,i} \}.
\end{align}
The $(L+1)$-tuple $\{R_l\}_{l\in [0:L]}$ will be called an admissible combination of coding rates for  $\{X_l\}_{l\in [0:L]}$ if, for $\epsilon \rightarrow 0$ and sufficiently large $n$, there exists a $(n,M_0,M_1,...,M_L)$-code for which $M_l\leq 2^{n(R_l+\epsilon)}, \forall l \in [0:L]$, and $p_{\max}<\epsilon$. The rate region, hereafter denoted as $\calR$, is the set of all admissible combinations of rates $\{R_l\}_{l\in [0:L]}$. 

Gelfand and Pinsker derived the rate region for the discrete lossless CEO problem under the CI condition\cite{gel1979}\footnote{The system model for the many-help-one problem is a special case of the system model for the lossless CEO problem investigated in \cite{gel1979} by Gelfand and Pinsker. In the CEO problem, the primary source is not encoded but rather observed from multiple helpers. Clearly, Gelfand and Pinsker's rate region is non-empty if and only if $H(X_0|X_1,\hdots,X_L)=0$ (this condition was referred to therein as ``completeness of observations"). From \cite{gel1979}, the rate region can be given by taking  into account the ``completeness of observations" for the encoding of the primary source~\cite{Oohama2008}.}. Below we summarize the Gelfand-Pinsker theorem when specialized to the lossless many-help-one problem.

\begin{theorem}
	\label{th:GelPin}
	(Gelfand and Pinsker \cite{gel1979}) The rate region is the convex closure of the set of all rates $\{R_l\}_{l\in [0:L]}$ satisfying the following conditions:
	\begin{enumerate}
		\item There exists an $L$-tuple $\{U_l\}_{l\in[L]}$ of discrete RVs taking values in $\calU_1 \times ... \times \calU_L$ such that $\{X_0,X_l,U_l\}_{l\in[L]}$ satifies the Markov chains $U_1...U_L \rightarrow X_1...X_L \rightarrow X_0$ and $U_l \rightarrow X_l \rightarrow X_{\bar{l}} \rightarrow U_{\bar{l}}, \forall l \in [L], l\neq \bar{l}$.
		\item $|\calU_l| \leq |\calX_l| +(L+1)2^{L-1}+1, \forall l \in [L]$.
		\item $R_0\geq H(X_0|U_1,...,U_L)$ and\\
			  $\sum_{l\in \calS} R_l \geq I\left(\{X_l\}_{l\in \calS};\{U_l\}_{l\in \calS}|\{U_l\}_{l\in \calS^\text{c}}\right),$ where\\ $\forall \calS \subseteq [L] \text{ and } \calS^\text{c}=[L]\backslash \calS $.
	\end{enumerate}
\end{theorem}
In \cite[Lemma 2.2]{Jana2009}, Jana showed that the computational complexity of Theorem~\ref{th:GelPin} can be reduced, i.e., the cardinality of the auxiliary RVs can be tightly bounded by $|\calU_l| \leq |\calX_l|, \forall l \in [L]$, for a broad class of multiterminal source coding problems, including the lossless many-help-one problem\footnote{The framework  provided by Jana includes the lossless many-help-one problem, when $(M,J,L)=(\text{any},1,0)$ and $S$ is deterministic. In \cite{Jana2009}, $M$ is the number of sources, $J$ is the number of sources which are reconstructed lossless, $L$ is the number of sources which are reconstructed within some distortion constraint, and $S$ is some side information.}. 

Even after the above reduction of cardinality, the optimization problem at hand remains highly complicated. Take, for instance, the case of $L=2$. To compute the lower convex boundary of $\calR$, we need to minimize the Lagrangian function
\begin{align}
	H(X_0|U_1,U_2)+&\mu_1 I(X_1;U_1|U_2) + \mu_2 I(X_2;U_2|U_1) \nonumber \\
\label{eq:sum_rate_1}
 + &\mu_3 I(X_1,X_2;U_1,U_2),
\end{align}
with $\mu_1,\mu_2,\mu_3>0$, over $p_{U_1|X_1}(u_1|x_1)$ and $p_{U_2|X_2}(u_2|x_2)$. Yet, the function in \eqref{eq:sum_rate_1} is  in general neither convex nor concave over  $p_{U_1|X_1}(u_1|x_1)$ and $p_{U_2|X_2}(u_2|x_2)$. For example, $H(X_0|U_1,U_2)$ is concave while $I(X_1;U_1|U_2)$ is convex over $p_{U_1|X_1}(u_1|x_1)$. Therefore, the optimization is surprisingly difficult even in the simplest case where all the sources are binary RVs.

\section{Binary Symmetric Case}

\begin{figure}[t]
	\centering
	\includegraphics[width=0.8\linewidth]{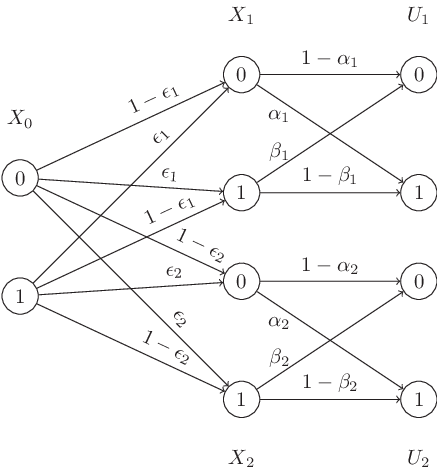}
	\caption{The binary symmetric case for $L=2$: the primary source and helpers are related via BSCs and the helpers and auxiliary RVs are related via BACs.}
	\label{fig:BinaryMarkov}
\end{figure}

In this section, we consider the case where the source $X_0$ is binary and uniformly distributed, i.e., $p_{X_0}(0)=p_{X_0}(1)=1/2$, and related to the helpers $X_l, l \in [L]$, via BSCs, i.e., $\epsilon_l\triangleq p_{X_l|X_0}(0|1)=p_{X_l|X_0}(1|0)$. Let us define the binary asymmetric channel (BAC) between the helpers and auxiliary RVs by the crossover probabilities $\alpha_l\triangleq p_{X_l|U_l}(1|0)$ and $\beta_l\triangleq p_{U_l|X_l}(0|1)$ for $l\in[L]$. Fig.~\ref{fig:BinaryMarkov} shows a schematic diagram of all RVs and the respective crossover probabilities for $L=2$. The optimization problem can be formulated as follows: for fixed $\epsilon_l$, determine all sets of $\{\alpha_l,\beta_l\}_{l \in [L]}$ which are on the lower convex boundary of the (optimal) rate region in Theorem~\ref{th:GelPin}. This optimization problem cannot be solved in closed form. Instead, a solution can be given within a target precision by means of an exhaustive numerical search.

Alternatively, driven by the results in \cite{Gu2007}, we derive an inner bound on the rate region's lower boundary based on the assumption of BSCs between the helpers and the auxiliary RVs, i.e., $\gamma_{l}=\alpha_l=\beta_l$ for $l \in [L]$. Importantly, later on we show by numerical examples that the proposed inner bound can be tight, mainly as the helpers turn out to be more degraded. In the following, we first examine the special case with two helpers and then extend our results to an arbitrary number of helpers.  

\subsection{Two Helpers}

\begin{theorem}\label{th:binary_2} If $(X_0,X_1,X_2)$ is a $3$-tuple of binary RVs and their joint pmf satisfies \eqref{eq:pmf_cond_dep}, with $p_{X_0}(0)=p_{X_0}(1)=1/2$, $p_{X_l|X_0}(0|1)=p_{X_l|X_0}(1|0)=\epsilon_l$, and $p_{U_l|X_l}(0|1)=p_{U_l|X_l}(1|0)=\gamma_l$ for some $0\leq \epsilon_l\leq1/2, l \in [2]$, then an inner bound on the rate region's lower boundary is given by
	\begin{align}
	\{(R_{0},R_{1},R_{2}): & \nonumber\\
		R_{0}= & h(\epsilon_1 \ast \gamma_1 )+h(\epsilon_2 \ast \gamma_2 ) -h(\epsilon_1 \ast \gamma_1 \ast \epsilon_2 \ast \gamma_2) , \nonumber \\
		R_{1}= & h(\epsilon_1 \ast \gamma_1 \ast \epsilon_2 \ast \gamma_2)  - h(\gamma_1), \nonumber  \\	
		R_{2}= & h(\epsilon_1 \ast \gamma_1 \ast \epsilon_2 \ast \gamma_2)  - h(\gamma_2) ,\nonumber \\
		R_{1}+R_{2}= & 1+h(\epsilon_1 \ast \gamma_1 \ast \epsilon_2 \ast \gamma_2)  - h(\gamma_1) - h(\gamma_2), \nonumber \\
		&(\gamma_1,\gamma_2) \in [0,0.5]^2 \}.
		\end{align}	
\end{theorem}
\begin{proof}
Given symmetric channels and uniformly distributed source symbols, the following holds:
	\begin{align}
	\label{eq:prop_1}
	H(X_0)=& H(X_l)=H(U_l)=1,\\
	\label{eq:prop_2}
	H(U_l|X_0)= &H(X_0|U_l), \\
	\label{eq:prop_3}
	H(U_1|U_2)= &H(U_2|U_1),
	\end{align}
	for $l \in [2]$. The information measures in \eqref{eq:sum_rate_1} can be reformulated as
	\begin{align}
	& H(X_0|U_1,U_2)=  H(X_0)+H(U_1|X_0)+H(U_2|X_0,U_1) \nonumber \\
	\label{eq:entropy_1}
	& \qquad - H(U_1) - H(U_2|U_1) \\
	\label{eq:entropy_2}
	& \quad=  H(U_1|X_0)+H(U_2|X_0) - H(U_2|U_1), \\
	\label{eq:mutual_info_1}
	& I(X_l;U_l|U_{\bar{l}})=  H(U_l|U_{\bar{l}})-H(U_l|X_l,U_{\bar{l}}) \\
	\label{eq:mutual_info_2}
	& \quad =  H(U_l|U_{\bar{l}})-H(U_l|X_l),\\
	\label{eq:mutual_info_3}
	& I(X_1,X_2;U_1,U_2)= H(U_1,U_2)-H(U_1,U_2|X_1,X_2) \\
	& \quad = H(U_1)+H(U_2|U_1) -H(U_1|X_1,X_2,U_2) \nonumber \\
	\label{eq:mutual_info_4}
	& \qquad-H(U_2|X_1,X_2) \\
	\label{eq:mutual_info_5}
	& \quad = H(U_1)+H(U_2|U_1)-H(U_1|X_1) -H(U_2|X_2),	
	\end{align}
	where $ l , \bar{l} \in [2]$, $ l \neq \bar{l} $. The steps can be justified as follows: in \eqref{eq:entropy_1} and \eqref{eq:mutual_info_4} we used the chain rule of entropy; \eqref{eq:entropy_2}, \eqref{eq:mutual_info_2}, and \eqref{eq:mutual_info_5} follow from from the Markov chain, i.e., $U_l \rightarrow X_l \rightarrow X_0 \rightarrow X_{\bar{l}} \rightarrow U_{\bar{l}}$, and the properties given in \eqref{eq:prop_1}. As shown by Wyner in \cite{Wyner1973_2}, the conditional entropy of two binary RVs $A$ and $B$ related by a BSC with crossover probability $\delta$ is given by $H(A|B)=h(\delta)$. For $U_l \rightarrow X_l \rightarrow X_0 $ and $U_1 \rightarrow X_1 \rightarrow X_0 \rightarrow X_2 \rightarrow U_2$, the end-to-end crossover probabilities are given by $\epsilon_l \ast \gamma_l$ and $\epsilon_1 \ast \gamma_1 \ast \epsilon_2 \ast \gamma_2$, respectively. Using this along with \eqref{eq:prop_2} and \eqref{eq:prop_3}, the conditional entropies in \eqref{eq:entropy_2}, \eqref{eq:mutual_info_2}, and \eqref{eq:mutual_info_5} can be given by
	\begin{align}
	\label{eq:entropy_3}
	&H(U_1|X_0)+H(U_2|X_0) - H(U_2|U_1) \nonumber \\ 
	& \qquad = h(\epsilon_1 \ast \gamma_1 )+h(\epsilon_2 \ast \gamma_2 ) - h(\epsilon_1 \ast \gamma_1 \ast \epsilon_2 \ast \gamma_2), \\
	&H(U_l|U_{\bar{l}})-H(U_l|X_l)= h(\epsilon_1 \ast \gamma_1 \ast \epsilon_2 \ast \gamma_2)  - h(\gamma_l),\\
	&H(U_2|U_1)-H(U_1|X_1)-H(U_2|X_2) \nonumber \\
	& \qquad = h(\epsilon_1 \ast \gamma_1 \ast \epsilon_2 \ast \gamma_2) - h(\gamma_1) - h(\gamma_2).
	\end{align}
The inner bound is then generated as the auxiliary parameters are ranged over $(\gamma_1,\gamma_2) \in [0,0.5]^2$. This completes the proof.
\end{proof}
\subsection{Extension to an Arbitrary Number of Helpers}
\begin{theorem}\label{th:binary_L} If $(X_0,X_1,...,X_L)$ is an $(L+1)$-tuple of binary RVs and their joint pmf satisfies \eqref{eq:pmf_cond_dep}, with $p_{X_0}(0)=p_{X_0}(1)=1/2$, $p_{X_l|X_0}(0|1)=p_{X_l|X_0}(1|0)=\epsilon_l$, and $p_{U_l|X_l}(0|1)=p_{U_l|X_l}(1|0)=\gamma_l$ for some $0\leq \epsilon_l\leq1/2, l \in [L]$, then an inner bound on the rate region's lower boundary is given by
	\begin{align}
	\{(R_{0},R_{1},...&,R_{L}): \nonumber \\
	R_{0}= &\sum\limits_{l \in [L]} h \left(\epsilon_{l} \ast \gamma_{l} \right) - \eta(\{\epsilon_l \ast \gamma_{l}\}_{l \in [L]}\}), \nonumber \\
\sum_{l\in \calS} R_l= &  \eta(\{\epsilon_l \ast \gamma_{l}\}_{l \in [L]}) - \eta(\{\epsilon_{l} \ast \gamma_{l}\}_{l \in \calS^\text{c}}) - \sum_{l \in \calS} h(\gamma_l),\nonumber \\
&  \forall \calS \subset [L] \text{ and } \calS^\text{c}=[L]\backslash \calS, \nonumber \\
\sum_{l\in [L]} R_l= & 1+  \eta(\{\epsilon_l \ast \gamma_{l}\}_{l \in [L]}) - \sum_{l \in [L]} h(\gamma_l), \nonumber \\
	&\{\gamma_l\}_{l\in[L]} \in [0,0.5]^L \},
	\end{align}
where $\eta(\{\cdot\})$ is defined in \eqref{eq:L_entropy}.  In particular, for $L=2$, $\eta ( \{ \epsilon_1 \ast \gamma_1 , \epsilon_2 \ast \gamma_2 \} ) = h ( \epsilon_1 \ast \gamma_1 \ast \epsilon_2 \ast \gamma_2 ) $. 		
\end{theorem}
\begin{proof}
The conditional entropy in Theorem~\ref{th:GelPin} can be given by
\begin{align}
\label{eq:multi_entropy_1}
& H\left(X_0|\{U_{l}\}_{l\in[L]}\right)	\nonumber \\
& \quad = H\left(X_0,\{U_{l}\}_{l\in[L]}\right) - H\left( \{U_{l}\}_{l\in[L]} \right) \\
\label{eq:multi_entropy_2}
& \quad =  \sum_{l \in [L]} H(U_{l} | X_0)  - H\left( \{U_{l}\}_{l\in[2:L]}|U_1 \right)\\
\label{eq:multi_entropy_3}
& \quad =   \sum\limits_{l \in [L]} h \left(\epsilon_{l} \ast \gamma_{l} \right) - \eta(\{\epsilon_l \ast \gamma_{l}\}_{l \in [L]}) 
\end{align}
for $L\geq 2$. The steps are justified as follows: (\ref{eq:multi_entropy_1}) is the chain rule of entropy; for (\ref{eq:multi_entropy_2}), the entropy of the primary source and the auxiliary RVs can be partitioned by the chain rule and simplified by the fact that $(U_1,...,U_L)$ are conditionally independent given $X_0$, i.e., $H(U_l|X_0,U_1,...,U_{l-1})=H(U_l|X_0)$, and the entropy of the auxiliary RVs can be reformulated by the chain rule; and (\ref{eq:multi_entropy_3}) follows from the properties in \eqref{eq:prop_1}-\eqref{eq:prop_3} for $l\in [L]$, the Markov chain, i.e., $H(U_l|X_0) = h(\epsilon_l \ast \gamma_l)$, and
\begin{align}
& H\left( \{U_{l}\}_{l\in[2:L]}|U_1 \right)   \nonumber \\
& \quad = -\sum_{u_1 \in \{0,1\}} p_{U_1}(u_1) \hspace{-0.3cm}  \sum_{\{u_l\}_{l \in [2:L]} \in \{0,1\}^{L-1}} \hspace{-0.7cm} p_{U_2...U_L|U_1} (u_2,...,u_l|u_1) \nonumber \\
\label{eq:multi_entropy_4}
& \qquad \times \ld p_{U_2...U_L|U_1} (u_2,...,u_l|u_1)  \\	
& \quad = -\sum_{\{u_l\}_{l \in [2:L]} \in \{0,1\}^{L-1}} \nonumber \\
& \qquad \times \Big( \sum_{x_0\in \{0,1\}} p_{U_2...U_L|X_0} (u_2,...,u_l|x_0) p_{X_0|U_1}(x_0|0) \Big)  \nonumber \\
\label{eq:multi_entropy_6}
& \qquad \times \ld \Big( \sum_{x_0\in \{0,1\}} p_{U_2...U_L|X_0} (u_2,...,u_l|x_0) p_{X_0|U_1}(x_0|0) \Big) \\
\label{eq:L_entropy}
& \quad \triangleq \eta(\{\epsilon_l \ast \gamma_{l}\}_{l \in [L]})
\end{align}
for $L\geq2$, with
\begin{align}
\label{eq:multi_entropy_5}
& p_{U_2...U_L|X_0} (u_2,...,u_l|x_0) = \prod_{l\in[2:L]} p_{U_l|X_0} (u_l|x_0) \\
\label{eq:pmf_ULX0}
& \quad = \prod_{l\in[2:L]}\left[\left(1-\epsilon_l\ast \gamma_l\right) \mathds{1}(u_l=x_0)+\left(\epsilon_l\ast \gamma_l\right) \mathds{1}(u_l\neq x_0)\right],
\intertext{and}
& p_{X_0|U_1} (x_0|0) = \big[\left(1-\epsilon_1\ast \gamma_1\right) \mathds{1}(x_0=0) \nonumber \\
\label{eq:pmf_U1X0}
& \qquad \qquad \qquad \quad +\left(\epsilon_1\ast \gamma_1\right) \mathds{1}(x_0\neq0)\big].
\end{align}
For $L<2$ we have $\eta(\{\epsilon_l \ast \gamma_{l}\}_{l \in [L]})=0$. In \eqref{eq:pmf_ULX0} and \eqref{eq:pmf_U1X0}, $\mathds{1}(\cdot)$ is the indicator function. The steps are justified as follows: \eqref{eq:multi_entropy_6} follows from the law of total probability, the Markov chain $U_1 \rightarrow X_0 \rightarrow U_2...U_L$, and the symmetric properties of the primary source and channels; and  \eqref{eq:multi_entropy_5} follows from the Markov chain $U_l \rightarrow X_0 \rightarrow U_{\bar{l}}$, for $l,\bar{l} \in [L]$ and $l \neq \bar{l}$.

\begin{figure}[t]
	\centering
	\begin{subfigure}{1\linewidth}
		\includegraphics[width=\linewidth]{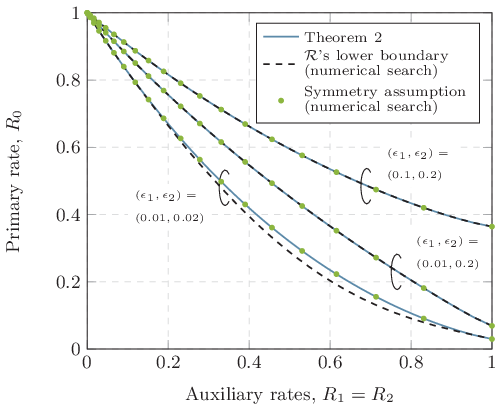}
		\caption{}
		\label{fig:ARR_sym}
	\end{subfigure}
	\begin{subfigure}{1\linewidth}
		\includegraphics[width=\linewidth]{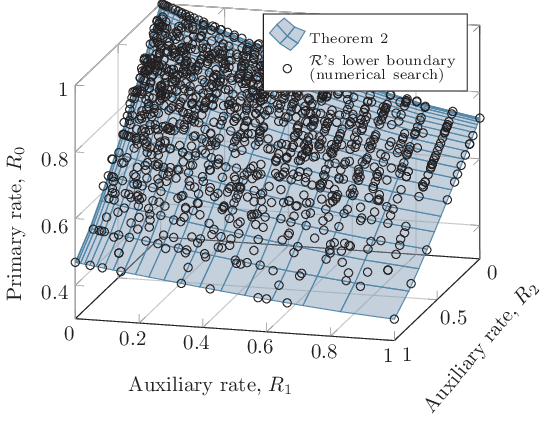}
		\caption{}
		\label{fig:ARR_3D}
	\end{subfigure}
	\caption{Rate region $\calR$: (a) symmetric rates $R_{1}=R_{2}\in[0,1]$ with $(\epsilon_1, \epsilon_2)\in\{(0.01,0.02),(0.01,0.2),(0.1,0.2)\}$, and (b) $(R_{1},R_{2})\in [0,1]^2$ with $(\epsilon_1, \epsilon_2)\in\{(0.1,0.2)\}$.}
	\label{fig:ARR_2}
\end{figure}

The conditional mutual information in Theorem~\ref{th:GelPin} can be given by
 \begin{align}
& I\left(\{X_l\}_{l\in \calS};\{U_l\}_{l\in \calS}|\{U_l\}_{l\in \calS^\text{c}}\right) \nonumber \\
 \label{eq:multi_mut_info_1}
 & \quad =  H\left(\{U_l\}_{l\in \calS}|\{U_l\}_{l\in \calS^\text{c}}\right)-   H\left(\{U_l\}_{l\in \calS}|\{X_l\}_{l\in \calS}\right) \\
 \label{eq:multi_mut_info_2}
& \quad = H\left(\{U_l\}_{l \in [L]}\right)- H\left(\{U_l\}_{l\in \calS^\text{c}}\right) - \sum_{l \in \calS} H(U_l|X_l)\\
& \quad =  \eta(\{\epsilon_l \ast \gamma_{l}\}_{l \in [L]}) - \eta(\{\epsilon_{l} \ast \gamma_{l}\}_{l \in \calS^\text{c}}) - \sum_{l \in \calS} h(\gamma_l),
\end{align}
$ \forall \calS \subset [L]\text{ and } \calS^\text{c}=[L]\backslash \calS$. The steps can be justified similarly as \eqref{eq:multi_entropy_1}-\eqref{eq:multi_entropy_3}. In particular, for $ \calS = [L] $, we have
 \begin{align}
& I\left(\{X_l\}_{l\in [L]};\{U_l\}_{l\in [L]}\right) \nonumber \\
\label{eq:multi_mut_info_3}
& \quad =   H\left(\{U_l\}_{l\in [L]}\right)-   H\left(\{U_l\}_{l\in [L]}|\{X_l\}_{l\in [L]}\right) \\
\label{eq:multi_mut_info_4}
& \quad = 1+  \eta(\{\epsilon_l \ast \gamma_{l}\}_{l \in [L]}) - \sum_{l \in [L]} h(\gamma_l).
\end{align}
The inner bound is then generated as the auxiliary parameters are ranged over $\{\gamma_l\}_{l\in[L]} \in [0,0.5]^L$. This completes the proof.  
\end{proof}

\section{Numerical Results and Discussion}
\label{NumResults}

In this section we illustrate our inner bounds in Theorem~\ref{th:binary_2} and Theorem~\ref{th:binary_L} by numerical examples. We show results for $L\in\{2,3\}$ with different values of $\epsilon_l$, $l \in [L]$.  We check these analytical bounds by performing an exhaustive numerical search (under the assumption that the auxiliary RVs are connected to the helpers through symmetric channels). Also, for comparison, we assess the exact rate region's lower boundary by performing an exhaustive search without any restriction on the symmetry, i.e., the helpers and auxiliary RVs being related via BACs with $\{\alpha_l,\beta_l\}_{l \in [L]}\in[0,0.5]^{2L}$.

In Fig.~\ref{fig:ARR_sym} we show our results in Theorem~\ref{th:binary_2} (blue line), numerical-search results for a symmetric channel between $X_l$ and $U_l$ (green dots), and the rate region's lower boundary (black dashed line). We show results for $L=2$ with $(\epsilon_1,\epsilon_2)=\{(0.01,0.02),(0.01,0.2), (0.1,0.2)\}$ and symmetric auxiliary rates $R_1=R_2=R$, i.e., $(R_0,R,R)$. The following can be observed: i) Theorem~\ref{th:binary_2} matches the simulation results for symmetric channels, and (ii) Theorem~\ref{th:binary_2} gives a tight inner bound on the rate region's lower boundary, especially for large values of $\epsilon_l$, i.e., as the helpers turn out to be more degraded versions of the primary source. Fig.~\ref{fig:ARR_3D} shows the rate $3$-tuples $(R_0,R_1,R_2)$ for $L=2$ and $(\epsilon_1,\epsilon_2)=\{(0.1,0.2)\}$. We show results of Theorem~\ref{th:binary_2} (blue plane) and  the rate region's lower boundary (black dots). The same conclusions as in Fig.~\ref{fig:ARR_sym} can be made by careful evaluation of a variety of setups.

\begin{figure}[t]
	\centering
		\includegraphics[width=\linewidth]{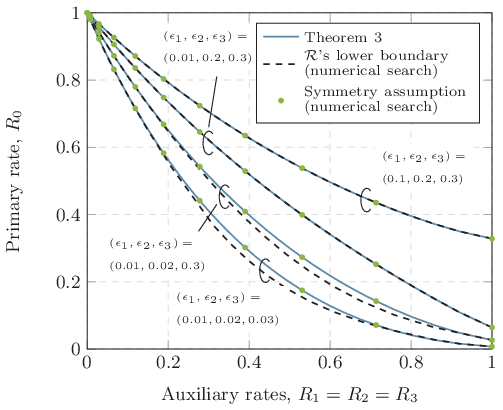}
	\caption{Rate region $\calR$: symmetric rates $R_{1}=R_{2}=R_{3}\in[0,1]$ with $(\epsilon_1, \epsilon_2,\epsilon_3)\in\{(0.01,0.02,0.03),$ $(0.01,0.02,0.3),$ $(0.01,0.2,0.3),$ $(0.1,0.2,0.3)\}$.}
	\label{fig:ARR_sym_3}
\end{figure}

In Fig.~\ref{fig:ARR_sym_3} we show our results in Theorem~\ref{th:binary_L} (blue line), numerical-search results for a symmetric channel between $X_l$ and $U_l$ (green dots), and the rate region's lower boundary  (black dashed line). We show results for $L=3$ with $(\epsilon_1, \epsilon_2,\epsilon_3)\in\{(0.01,0.02,0.03),$ $(0.01,0.02,0.3),$ $(0.01,0.2,0.3),$ $(0.1,0.2,0.3)\}$ and symmetric auxiliary rates $R_1=R_2=R_3=R$, i.e., $(R_0,R,R,R)$. The same conclusions as for $L=2$ can be drawn.

 \section*{Acknowledgment}
 This work was supported in part by the Federal Ministry of Education and Research within the programme ``Twenty20- Partnership for Innovation" under contract 03ZZ0505B - ``fast wireless", in part by the German Research Foundation (DFG) within the SFB 912 ``Highly Adaptive Energy-Efficient Computing (HAEC)", and in part by the São Paulo Research Foundation (FAPESP) in Grant 2016/05847-0.

\bibliographystyle{IEEEtran}
\bibliography{refs}

\end{document}